\documentclass[preprint,aps]{revtex4}
\usepackage{epsfig}
\begin{document}
\title{ON THE DIRAC OSCILLATOR}
\author{R. de Lima Rodrigues\\
Unidade Acad\^emica de Educa\c{c}\~ao\\ Universidade Federal de
Campina Grande, Cuit\'e - PB, CEP 58.175-000- Brazil
\\ Centro Brasileiro de Pesquisas F\'\i sicas (CBPF)\\
Rua Dr. Xavier Sigaud, 150, CEP 22290-180, Rio de Janeiro, RJ,
Brazil }

\begin{abstract}
In the present work we obtain a new representation for the Dirac
oscillator based on the Clifford algebra $C\ell_7.$ The symmetry
breaking and the energy eigenvalues for our model of the Dirac
oscillator are studied in the non-relativistic limit.

\vspace{1cm} PACS numbers: 11.30.Pb, 03.65.Fd, 11.10.EF

\vspace{5cm} E-mail to RLR is rafael@df.ufcg.edu.br or
rafaelr@cbpf.br.

\vspace{2cm} To appear in Physics Letters A.
\end{abstract}

\maketitle

\newpage

\section{Introduction}

The relativistic tridimensional isotropic harmonic oscillator has
been introduced many years ago by It\^o, Mori and Carriere
\cite{ITO}, with the Dirac Hamiltonian linear in the position $\vec
r$ and momentum $\vec p,$ with the replacement of $\vec p$ by $\vec
p - im\omega\beta\vec r,$ where $i=\sqrt{-1}, m$ the mass and
$\omega$ the oscillator frequency. This system is an exactly soluble
model which has unusual accidental degeneracies in its spectrum
\cite{Cook}.

The system analyzed in \cite{ITO}, was denominated Dirac oscillator
by Moshinsky and Szczepaniak \cite{Mosh89}. The Dirac oscillator has
been investigated in several contexts
\cite{MZ,DB,Quesne90,Beni90,Lang91,Castanos91,Dixit92,villalba94,martinez95,SG01,carlos03,alhai04,joseph04,ant04,QT05,mustafa07,RJV06NE}.
The Dirac oscillator with a generalized interaction was treated by
Casta\~nos {\it et al.} \cite{Castanos91}.  Dixit {\it et al.}
\cite{Dixit92} have obtained a parity invariant Dirac oscillator
with scalar coupling by doubling the number of components and  using
a representation of the Clifford algebra $C\ell_7$. These works
motivate the construction of a new linear Hamiltonian in terms of
the momentum, position and mass coordinates, through a set of seven
mutually anticommuting $8\times8$-matrices yielding a representation
of the Clifford algebra $C\ell_7.$

In the present work we study a new formulation of the Dirac
oscillator using the Clifford algebra $C\ell_7$ which,  in the
non-relativistic limit leads to the 3D isotropic oscillator with a
correction term for both signs of energy. The correction term is
different from those in the other formulations and will be
interpreted in the following.

\section{Generalized Dirac Oscillator}

 The  Clifford algebra $C\ell_7$ is defined by a set of 7 objects
satisfying the anticommutation relations

\begin{equation}
\label{E1} [\Gamma_a, \Gamma_b]_+=2\delta_{ab}{\bf 1}, \quad a, b
=1, 2, \cdots 7.
\end{equation}
The irreducible representations of $\Gamma_a$ are provided by the $8\times8$
matrices given by

\begin{equation}
\label{E2} \vec\Gamma={\bf 1}_{2\times2}\otimes\vec\alpha, \quad
\Gamma_4={\bf 1}_{2\times2}\otimes\beta, \quad
\Gamma_{4+i}={\tilde\Gamma}_i=\rho_i\otimes\gamma_5,
\end{equation}
where $\rho_i, i=1, 2, 3,$ are a set of the Pauli matrices and

\begin{equation}
\label{E3} \vec\alpha=\tau_1\otimes\vec\sigma, \quad (i=1, 2, 3),
\quad \beta=\tau_3\otimes{\bf 1}_{2\times2}, \quad
\gamma_5=\alpha_1\alpha_2\alpha_3\beta=\tau_2\otimes{\bf
1}_{2\times2}.
\end{equation}
Here, $\rho_i, \tau_i$ and $\sigma_i$ are three sets of the Pauli
matrices which act in different space. Now, we build the Dirac
oscillator Hamiltonian linear in the position $\vec r$, momentum
$\vec p$ and mass $M$ as:

\begin{eqnarray}
\label{E6a}
H &&= c\vec\Gamma\cdot\vec p +{\Gamma_4}Mc^2
+cM\omega\vec{\tilde\Gamma}\cdot\vec r.
\end{eqnarray}
The above Hamiltonian gives

\begin{equation}
\label{E6b}
H^2= c^2{\vec p}^2 +M^2c^4 +c^2M^2\omega^2r^2-
i\hbar c^2 M\omega \vec\Gamma\cdot\vec{\tilde\Gamma}.
\end{equation}

To interpret the last term in $H^2$
 we analyze the structure of the total angular momentum
associated with the Hamiltonian $H.$

It is easy to verify the commutation relations:

\begin{equation}
\label{E8} [H, L_i]_-= -i\hbar c(\vec\Gamma\wedge\vec p)_i -i\hbar
cM\omega(\vec{\tilde\Gamma}\wedge\vec r)_i
\end{equation}

\begin{equation}
\label{E10} [H, S_i]_-= i\hbar c(\vec\Gamma\wedge\vec p)_i
\end{equation}
where $\vec L={\bf 1}_{8\times8}\otimes \vec r\wedge\vec p$ and $
\vec S= -\frac{i\hbar}{4} {\bf 1}_{2\times2}\otimes
(\vec\alpha\wedge\vec\alpha).$

Thus

\begin{equation}
\label{E11}
[H, L_i+S_i]_-= -i\hbar
cM\omega(\vec{\tilde\Gamma}\wedge\vec r)_i\neq 0, \quad i=1, 2, 3.
\end{equation}

Now we compute the commutator of $H$ with another
spin like operator $\vec I,$
 which we define as being

\begin{eqnarray}
\label{E12} \vec I&&= -i\frac{\hbar}{4}
\vec{\tilde\Gamma}\wedge\vec{\tilde\Gamma}
\end{eqnarray}
with

\begin{equation}
\label{E13}
[I_i, I_j]_-= i\hbar\epsilon_{ijk}I_k, \quad (i, j, k=1, 2, 3).
\end{equation}
We obtain

\begin{equation}
\label{E14} [H, I_i]_-= i\hbar cM\omega(\vec{\tilde\Gamma}\wedge\vec
r)_i, \quad i=1, 2, 3.
\end{equation}

Thus we see that the operator
\begin{eqnarray}
\label{E15}
\vec J &&= \vec L + \vec S +\vec I
\end{eqnarray}
with
\begin{equation}
\label{E13a} [J_i, J_j]_- =i\hbar\epsilon_{ijk}J_k  \quad
i,j,k=1,2,3,
\end{equation}
satisfies the equation

\begin{equation}
\label{E15a}
[H, \vec J]_-= [H, \vec L +\vec S+\vec I]_-=0.
\end{equation}

Thus we may identify $\vec J$ as the total conserved angular
momentum. When $\omega =0,$ $\vec J_D=\vec L +\vec S$ commutes also
with the Hamiltonian in equation (\ref{E8}), i.e.

\begin{equation}
\label{E17}
[H(\omega=0), \vec L +\vec S]_-=0.
\end{equation}
Note that the operator $H(\omega=0)$ is a direct sum of two
Hamiltonians of the free Dirac particle, viz.,

\begin{equation}
\label{E18}
H(\omega=0)= c\vec\Gamma\cdot\vec p +
Mc^2\Gamma_4
=\left(
\begin{array}{cc}
c\vec\alpha\cdot\vec p +
Mc^2\beta & 0\\
0 & c\vec\alpha\cdot\vec p +
Mc^2\beta
\end{array}
\right).
\end{equation}
In this case with $\omega=0,$ the operator $\vec I $
commutes also with $H(\omega=0):$

\begin{equation}
\label{E19}
[H(\omega=0), I_i]_-=0, \quad (i=1, 2, 3).
\end{equation}

Therefore, the operators $I_i,$ commuting with $H(\omega=0),$
generate a global symmetry of $SU(2)$ between the Dirac particles
described by the Hamiltonians in the lower and upper sectors.
 The doublet of fermionic particles described by this Hamiltonian
can be labeled by  value 1/2 of the I-spin, and the eigenvalues of
$I_3=\frac{\hbar}{2}$ and  $I_3=-\frac{\hbar}{2}.$ From equation
(\ref{E12}), we obtain:

\begin{equation}
\label{E20} I^2= \frac{3}{4}\hbar^2\left(
\begin{array}{cc}
{\bf 1}_{4\times4}& 0\\
0 & {\bf 1}_{4\times4}
\end{array}
\right), \quad I_3=\frac{\hbar}{2}\left(
\begin{array}{cc}
{\bf 1}_{4\times4}& 0\\
0 & -{\bf 1}_{4\times4}
\end{array}
\right).
\end{equation}
The interaction term in equation (\ref{E6a}) is dependent on the
operators of the ordinary spin  $S_i$ and of the I-spin $I_i$.
Indeed, using the definitions given by equations (\ref{E3}),
(\ref{E8}) and (\ref{E12}) we can write the forms of $H$ and $H^2,$
respectively, as

\begin{eqnarray}
\label{EH1} H &&= {\bf 1}_{2\times 2}\otimes(c\vec\alpha\cdot\vec p
+\beta Mc^2) +\frac{2}{\hbar}cM\omega\vec I\cdot\vec r\nonumber\\
H^2 &&= c^2{\vec p}^2 +M^2c^4 +c^2M^2\omega^2r^2+ \frac{4}{\hbar}
c^2M\omega \vec S\cdot \vec I.
\end{eqnarray}

At this stage, we can justify that the matrices of I-spin (1/2 in
our case), represent an inner symmetry of the doublet of free Dirac
particles, given by above expression for the total angular momentum.
Indeed, we notice that $\vec S +\vec I$ is the true  total spin of
the Dirac oscillator described by the Hamiltonian (\ref{EH1}).

 Next, we consider the solution of our model of the Dirac oscillator in the
non-relativistic limit. If we decompose the eigenfunction $\Phi$ of
$H$ with eigenvalue $E_R$ in the form

\begin{equation}
\Phi= \left[\matrix{v_{1} \cr w_{1} \cr v_{2} \cr w_{2}}\right],
\end{equation}
where $v_{1}, w_{1}, v_{2}$ and $w_{2}$ are two-component spinors,
the eigenvalue equation

\begin{equation}
\label{EH2} H\Phi  = \{{\bf 1}_{2\times2}\otimes (c\vec{\alpha
}\cdot\vec{p} + \beta Mc^{2}) + {2\over \hbar }cM\omega (\vec
I\cdot\vec r)\}\Phi = E_{R}\Phi
\end{equation}
gives:

\begin{eqnarray}
\label{generallabel lc}
{}&&\pmatrix{c\tau_{1}\otimes\vec{\sigma}\cdot\vec{p}+\tau_{3}\otimes{\bf
1}_{2\times2}Mc^{2}&0 \cr 0& c\tau_{1}\otimes\vec{\sigma
}\cdot\vec{p}+\tau_{3}\otimes{\bf
1}_{2\times2}Mc^{2}}\left[\matrix{v_{1} \cr
w_{1} \cr v_{2} \cr w_{2}}\right]\nonumber\\
{}&&+cM\omega (\vec{\rho }\cdot \vec{r}){\otimes{\pmatrix{0&-i{\bf
1}_{2\times2} \cr i{\bf 1}_{2\times2}&0}}}\left[\matrix{v_{1} \cr
w_{1} \cr v_{2} \cr w_{2}}\right]= E_{R}\left[\matrix{v_{1} \cr
w_{1} \cr v_{2} \cr w_{2}}\right],
\end{eqnarray}
where

\begin{equation}
\vec\rho\cdot\vec{r}= \left(\matrix{x_{3}&r_{-} \cr
r_{+}&-x_{3}}\right), \quad r_{\mp }\equiv  x_{1}\mp ix_{2}.
\end{equation}
Thus we get for the spinors  $v_{1}, w_{1}, v_{2}$ and $w_{2}$, the
following relations:

\begin{eqnarray}
E_{R}v_{1}&&= Mc^{2}v_{1}+ c\vec{\sigma }\cdot\vec{p}w_{1}+ cM\omega
(-ix_{3}w_{1}-ir_-w_{2}),
\nonumber\\
E_{R}w_{1}&&= -Mc^{2}w_{1}+ c\vec{\sigma }\cdot\vec{p}v_{1}+ cM\omega
(ix_{3}v_{1}+ir_{-}v_{2}),
\nonumber\\
E_{R}v_{2}&&= Mc^{2}v_{2}+ c\vec{\sigma }\cdot\vec{p}w_{2}+ cM\omega
(-ir_{+}w_{1}+ix_{3}w_{2}),
\nonumber\\
E_{R}w_{2}&&= -Mc^{2}w_{2}+ c\vec{\sigma }\cdot\vec{p}v_{2}+ cM\omega
(ir_{+}v_{1}-ix_{3}v_{2}),
\end{eqnarray}
 which show that in the non-relativistic limit,
for $E_{R}\rightarrow E_+^{\prime}+Mc^{2}$ (positive energy), the
components $v_{1}$ and $v_{2}$ are predominant and $w_{1}\rightarrow
(v_{1}/c)\rightarrow 0$ and $w_{2}\rightarrow (v_{2}/c)\rightarrow
0.$ On the other hand, for $E_{R}\rightarrow -E_-^{\prime}-Mc^{2}$
(negative energy), the components $w_{1}$ and $w_{2}$ are
predominant and $v_{1}\rightarrow  (w_{1}/c)\rightarrow 0$ and
$v_{2}\rightarrow  (w_{2}/c)\rightarrow 0.$

Now, the eigenvalue equation for $H^{2}$  can be
simplified to give

\begin{equation}
H^{2}\Phi  = E^{2}_{R} \left[\matrix{v_{1} \cr w_{1} \cr v_{2} \cr
w_{2}}\right]= \left\{c^{2}p^{2}+ M^{2}c^{4}+ c^{2}M^{2}\omega
^{2}r^{2}+Mc^{2}\hbar\omega \vec{\rho } \otimes \tau_{3}\otimes
\cdot\vec{\sigma }\right\} \left[\matrix{v_{1} \cr w_{1} \cr v_{2}
\cr w_{2}}\right],
\end{equation}
or

\begin{equation}
\label{E26}
{E^{2}_{R} - M^2c^{4}\over 2Mc^{2}}\left[\matrix{v_{1} \cr w_{1}
\cr v_{2} \cr w_{2}}\right]= \left\{{p^{2}\over 2M} +
{1\over 2}M\omega r^{2}+ \hbar \omega \pmatrix{{1\over2}\sigma_{3}&0&
\sigma _{-}&0 \cr 0&-{1\over2}\sigma_{3}&0&-\sigma _{-} \cr
\sigma_{+}&0&-{1\over2}\sigma_{3}&0 \cr 0&-\sigma_{+}&0&{1\over2}\sigma_{3}}
\right\}\left[\matrix{v_{1} \cr w_{1} \cr v_{2} \cr w_{2}}\right],
\end{equation}
where

\begin{equation}
{1\over 2}\vec{\rho }\otimes \tau_{3}\otimes \cdot\vec{\sigma } =
\pmatrix{{1\over2}\sigma_{3}&0&\sigma _{-}&0 \cr 0&-{1\over2}
\sigma_{3}&0&-\sigma _{-} \cr \sigma_{+}&0&-{1\over2}\sigma_{3}&0
\cr 0&-\sigma_{+}&0&{1\over2}\sigma_{3}}, \quad\sigma _{\pm} \equiv
{1\over 2}(\sigma _{1}\pm i\sigma _{2}).
\end{equation}
Observing that equation (\ref{E26}) gives us coupled relations only
between $(v_{1}, v_{2})$ or $(w_{1}, w_{2})$, we have:

\begin{eqnarray}
\label{E28}
{E^{2}_{R} - M^2c^{4}\over 2Mc^{2}}\pmatrix{v_{1} \cr v_{2}} &&=
\left\{{p^{2}\over 2M} + {1\over 2}M\omega r^{2}+ \hbar\omega
\pmatrix{{1\over2}\sigma_{3}&\sigma _{-} \cr \sigma_{+}&-{1\over 2}
\sigma _{3}}\right\}
\pmatrix{v_{1} \cr v_{2}},
\nonumber\\
{E^{2}_{R} - M^2c^{4}\over 2Mc^{2}}\pmatrix{w_{1} \cr w_{2}}&&=
\left\{{p^{2}\over 2M} + {1\over 2}M\omega r^{2}- \hbar\omega
\pmatrix{{1\over2}\sigma_{3}&\sigma_{-} \cr \sigma_{+}&-{1\over 2}
\sigma_{3}}\right\}\pmatrix{w_{1} \cr w_{2}},
\end{eqnarray}
where

\begin{equation}
\pmatrix{{1\over2}\sigma_{3}&\sigma _{-} \cr \sigma_{+}&-{1\over 2}
\sigma _{3}}= {1\over 2}\{\vec{\rho }\otimes {\bf
1}_{2\times2}\}\cdot \{{\bf 1}_{2\times2}\otimes \sigma \}= {1\over
2}\vec{\rho }\otimes\cdot\vec{\sigma }.
\end{equation}
Next, using the fact that  $v_{1}$ and $v_{2}$ are large compared to
$w_{1}$ and $w_{2}$ in the case $E_{R}\rightarrow
E^{\prime}_{+}+Mc^{2}\Rightarrow E^{2}_{R} - M^2c^{4}\rightarrow
2Mc^{2}E^{\prime}_{+},$ and $E_{R}\rightarrow
-E^{\prime}_{-}-Mc^{2}\Rightarrow E^{2}_{R} - M^2c^{4}\rightarrow
-2Mc^{2}E^{\prime}_{-},$ equations (\ref{E28}) give us

\begin{eqnarray}
\label{HNR} E^{\prime}_{+}\pmatrix{v_{1} \cr v_{2}} &&=
\left({{p^{2}\over 2M} + {1\over 2}M\omega r^{2}+ {1\over
2}\hbar\omega\vec{\rho}\otimes\cdot\vec{\sigma }}
\right)\pmatrix{v_{1} \cr v_{2}},
\nonumber\\
E^{\prime}_{-}\pmatrix{w_{1} \cr w_{2}} &&= \left({{p^{2}\over 2M} +
{1\over 2}M\omega r^{2}- {1\over 2}\hbar
\omega\vec{\rho}\otimes\cdot \vec{\sigma }}\right)\pmatrix{w_{1} \cr
w_{2}},
\end{eqnarray}
where $E^{\prime}_{\pm}$ are assumed to be small in comparison with
$Mc^2.$  The operator $\vec{\rho}\otimes\cdot\vec{\sigma }$ commutes
with all the other terms of these two Hamiltonians, so that we can
substitute it by the eigenvalues 1 and -3,  when acting
 on the triplet states,
$V_{T}$ and $W_{T},$ and on the
 singlet states, $V_{S}$ and $W_{S},$ respectively. Hence we get:

\begin{eqnarray}
E^{\prime}_{+}V_{T} &&= \left({{p^{2}\over 2M} + {1\over 2}M\omega
r^{2}+ {1\over 2}\hbar \omega}\right)V_{T}, \quad
E^{\prime}_{-}W_{T} = \left({{p^{2}\over 2M} + {1\over 2}M\omega
r^{2}- {1\over 2}\hbar \omega }\right)W_{T},
\nonumber\\
E^{\prime}_{+}V_{S} &&= \left({p^2\over 2M} + {1\over 2}M\omega
r^{2}- {3\over 2}\hbar \omega \right)V_{S},  \quad
E^{\prime}_{-}W_{S} = \left({{p^{2}\over 2M} + {1\over 2}M\omega
r^{2}+ {3\over 2}\hbar \omega }\right)W_{S}.
\end{eqnarray}

The energy spectra of these
 Hamiltonians are then given by:

\begin{eqnarray}
\left[E^{\prime}_{+}\right]^{(n)}_{T}&&= (n+2)\hbar \omega, \quad
\left[E^{\prime}_{-}\right]^{(n)}_{T}= (n+1)\hbar \omega, \quad
\left[E^{\prime}_{+}\right]^{(n)}_{S}= n\hbar \omega
\nonumber\\
\left[E^{\prime}_{-}\right]^{(n)}_{S}&&= (n+3)\hbar \omega, \quad
(n=0,1,2 \cdots), n=\ell+2m,
\end{eqnarray}
 which reveal an asymmetry between the  positive and negative energy spectra.
 For example, when $\left[E^{\prime}_{+}\right]^{(0)}_{S}=0,$
we have $E_{R}=\left[E^{\prime}_{+}\right]^{(0)}+ Mc^{2}=Mc^{2};$ on
the other hand, when $ \left[E^{\prime}_{-}\right]^{(0)}_{S}=3\hbar
\omega, E_{R}=-Mc^{2}$ is absent, since
$E_{R}=-\left[E^{\prime}_{-}\right]^{(0)}_{S}- Mc^{2}=
-Mc^{2}-3\hbar\omega. $ The interesting question about the existence
or not of an interaction that inverts this asymmetry of the spectra
derived above, in  the non-relativistic limit, can be responded in
the affirmative. This interaction corresponds to changing the sign
of $\omega$ in equations (\ref{E6a}) and (\ref{E6b}). The
nonequivalence of the spectra in these two cases follows from the
nonexistence of a unitary transformation satisfying the following
conditions: $\Gamma_{i}\rightarrow \Gamma_{i}, \quad
\Gamma_4\rightarrow \Gamma_4$ and $\tilde\Gamma_i\rightarrow
-\tilde\Gamma_i$, i.e., $\vec{\alpha}\rightarrow \vec{\alpha}, \quad
\beta\rightarrow\beta, \quad\gamma_{5}\rightarrow \gamma_{5}$ and
$\vec{\rho}\rightarrow -\vec{\rho }$, since the representations
$\vec{\rho}$ and $-\vec{\rho}$ are inequivalent.

\section{Conclusion}

We have found a new representation for a Dirac oscillator via the
Clifford algebra $C\ell_7.$ With the introduction of the interaction
dependent on the $I$-spin in (\ref{E12}), for $\omega\neq 0,$ the
global symmetry $SU(2)$ that exists in the case $\omega=0$ is broken
and the $I$-spin degrees of freedom convert to the degrees of
freedom of spin and orbital angular momentum, according to equation
(\ref{E15}).

In the context of a
gauge field theory with the local symmetry of $SU(2)$ spontaneously broken,
 such a phenomenon of convention of the degrees of
freedom $I$-spin to spin \cite{Gauge} occurs. However, the  breaking
of the  global symmetry $SU(2)$, as in our Dirac oscillator model,
has not been investigated in the literature. Our Dirac oscillator
model is not manifestly covariant, however, it is
quantum-mechanically well-defined. Interestingly, the Hamiltonian
does not commute with the ordinary angular momentum operator $\vec
J=\vec L+\vec S,$ but a new $I$-spin must be added.

The symmetry breaking brought out here for the Dirac
oscillator was studied in the
non-relativistic limit
 when the additional constraint provided by the Dirac
equation is fully implemented.

The formulation of the
Dirac oscillator Hamiltonian \cite{Mosh89} in terms of the Wigner ladder operators
\cite{JR90}
permits a purely algebraic treatment for the relativistic problem \cite{RV02},
the details of which will be published separately.

 Let us conclude with a discussion on the relationship between the
 new Dirac oscillator and other proposals including the $4\times4$
 oscillator with vector coupling and the $8\times8$ oscillator scalar coupling.
 The usual Dirac oscillator Hamiltonian in the non-relativistic limit  leads to
that of a 3-dimensional isotropic oscillator shifted by a constant
term plus a ${\vec L}\cdot {\vec S}$ coupling term  for both signs
of energy. In another work, Dixit {\it et al.} \cite{Dixit92} have
considered  the Dirac oscillator with scalar coupling which in the
non-relativistic limit leads to a harmonic oscillator Hamiltonian
plus a $\vec\sigma\cdot \hat r$ coupling term, where $\hat
r=\frac{\vec r}{r}.$
 In the new Dirac oscillator presented in this paper the correction term is different.
Indeed, our Dirac oscillator Hamiltonian in the non-relativistic
limit  leads to that of a 3-dimensional isotropic oscillator plus a
$\vec{\rho}\otimes\cdot\vec{\sigma }$ coupling term.

\centerline{\bf Aknowledgments}

 The author was supported in part by CNPq (Brazilian Research Agency).
He wishes to thank J. A. Helayel Neto for the kind of hospitality at
CBPF-MCT where this work was mostly carried out. He also thanks the
staff of the CBPF and UFCG. This work begun in collaboration with
Jambunatha Jayaraman (In memory) and Arvind Narayan Vaidya (In
memory), author's Ph. D. supervisors, whose advises and
encouragement were fundamental in the development of the thesis.

\end{document}